\documentclass[aps,pra,reprint,superscriptaddress,nobibnotes]{revtex4-2}
\usepackage{amsmath,amsfonts,amssymb}
\usepackage{braket}
\usepackage{ulem} 
\usepackage[dvipsnames]{xcolor}
\usepackage{comment}
\usepackage{graphicx}
\usepackage{xcolor}
\usepackage{dsfont}
\usepackage{epstopdf}
\usepackage{tikz}
\usepackage{subcaption}
\usepackage{siunitx}
\usepackage{amsthm}
\usepackage{yfonts}
\usepackage[colorlinks=true,linkcolor=blue,citecolor=blue,urlcolor=blue]{hyperref}

\newcommand{\pro}[1]{\vert #1\rangle\langle  #1\vert}

\usepackage{quantikz}
\usepackage{bm}

\begin{document}
\title{Solving a Linear System of Equations on a Quantum Computer by Measurement}

\author{Alain Giresse Tene}
\affiliation{ University of Kwazulu-Natal, Private Bag X54001, Durban 4000, South Africa}
\affiliation{UKZN Centre for Quantum Computing and Technology, Durban, South Africa}
\author{ Thomas Konrad}
\affiliation{ University of Kwazulu-Natal, Private Bag X54001, Durban 4000, South Africa}
\affiliation{UKZN Centre for Quantum Computing and Technology, Durban, South Africa}

\begin{abstract}
We present a variational algorithm for fault tolerant quantum computing to solve a system of linear equations  which directly maximises the parameters of the target fidelity. This so-called measurement test algorithm can be applied to any computational task with a solution that is represented as eigenvector of a self-adjoint matrix. 
The solution is prepared as state of a register in the quantum computer by a von Neumann measurement of a corresponding observable, which is implemented using the phase estimation algorithm. The probability to project the system thus into the unknown target state, which equals the target fidelity,  
is measured in terms of relative frequencies and iteratively optimised to read out the target state.   
The new algorithm overcomes three issues of previous variational quantum algorithms: i) It does not rely on a decomposition in terms of Pauli strings and therefore can compute eigenvectors of dense matrices. ii) The accuracy  is not limited by the condition number $\kappa$ of the matrix, provided a logarithmic number ($O(\log\kappa)$) of qubits is used to encode the eigenvalues and iii) the target fidelity $F_T = 1-\epsilon$ can be reached with an accuracy $\epsilon$ that scales with $1/N$ for $N$ measurements per iteration. We demonstrate this by numerical simulations for dense random real-valued $16\times 16$ matrices with non-vanishing determinant.      
\end{abstract}

\maketitle
\section{Introduction}
Quantum computers process quantum information by transforming states of quantum systems. A basic unit of quantum information, a quantum bit or qubit,  can be stored in a quantum system with a two dimensional state space, e.g.,  an atom with two relevant internal energy levels or a photon which can propagate only along two paths. A qubit can thus be characterised by a normalised vector with two complex components (or three real numbers).  A register of a quantum computer with $d$ qubits has the capacity to store an exponential number, $L= 2^n$, of components  and process them simultaneously by a physical transformation of the state of the register. This can be exploited by encoding information in the amplitudes $r_i$ or phases $\phi_i$ of the complex components $c_i=r_i \exp(i\phi_i)$ of the state of the quantum register 
\begin{equation}
\ket{\psi}= \sum_{i=0}^{2^n-1} c_i\ket{i}
\end{equation}   
where $\ket{i}\equiv\ket{i_{n-1}, \ldots i_1, i_0}$ is the $i$th basis state in binary representation with respect to the basis states  $i_j=0,1$ of $n$ qubits, $j=0\ldots n-1$.  
A quantum computation with amplitude encoding prepares a state with amplitudes $r_i$ that  represent the solution to a computational task. However, the number of steps to detect the $L$ amplitudes by measurements of the computational basis would scale exponentially with the number $n$ of qubits.     

For example, it has been suggested to compute the solutions of linear systems of equations by matrix inversion in a quantum computer \cite{HHL2009} and to find the solution of optimisation tasks using a quantum gradient descent method for optimisation problems \cite{Rebentrost_2019}. Although amplitude-encoding methods prepare certain target state solutions with exponential speed up compared to their classical counterparts, a direct detection of the amplitudes would compensate the speed up.     
 
Variational quantum algorithms  (VQAs) \cite{Cerezoetal21} use a specific technique to approximate the desired target state more efficiently than a full state tomography which requires a multitude of measurements \cite{Bharti.et.al2022}. They prepare an input state depending on a set of parameters (a so-called ansatz), measure the value of the objective function on a quantum computer, and optimise it by varying the input parameters based on classical computation. Once the objective function reaches a value sufficiently close to the optimal one, the solution of the computational task can be read off from the corresponding input state, which  is determined by the classical input parameters. 

VQAs serve, for example, to compute energies and electronic configurations of molecules and condensed matter systems \cite{OBrian.et.al2014, Grimsley.et.al19, Tilly.et.al22, Yashioka.et.al22, Innan.et.al23,Li.et.al25}, to determine nuclear structure \cite{Romero.et.al22}, to find the solution of combinatorial optimisation problems \cite{Farhi.et.al2014, Moll.et.al2018,Diez-Valle.et.al2023}, to solve certain partial differential equations \cite{Childs2021,Balducci2022,Jin2023,Nguyen2025}, including non-linear Schr\"odinger equations by means of multiple state copies \cite{Lubasch.et.al20},  and might offer a quantum advantage when searching for the solution of non-polynomial classification problems \cite{Jaeger.et.al2023}. 

The evaluation of the objective function can be implemented using a technique (the Hadamard test \cite{Nielsen.Chuang00}) that requires measuring various observables given by products of Pauli operators (so-called Pauli strings) which, in general, do not commute and therefore involve non-vanishing standard deviations due to Heisenberg's uncertainty relation. Moreover, the measurement of each Pauli string requires the execution of a particular quantum circuit with a multitude of runs. Since the number of Pauli strings of generic observables is expected to grow exponentially with the number of qubits in the input register, the number of quantum circuit executions would render VQAs to compute the expectation value of arbitrary observables inefficient. For this reason, so far only computational tasks corresponding to sparse matrices involving a few Pauli strings are considered to be solvable using  VQAs.

We present here the measurement test algorithm applied to solve a linear system of equations with unique solution and a  dense (non-sparse) matrix. In Section \ref{section2} we study the algorithm and  analyse in Section \ref{section3} the accuracy of its results. We summarise and discuss our findings  in Section \ref{section4}. 

\section{The Measurement Test Algorithm }\label{section2} 
 Here the aim of the new algorithm is to find the solution of the linear system of equations
\begin{equation}
M \mathbf{x} = \mathbf{b}\,,   
\label{eqnM}
\end{equation}
 where $M$ is an $L \times L$ matrix with real entries and $\mathbf{x}, \mathbf{b} \in \mathbb{R}^L$. The determinant of $M$ does not vanish, thus there is exactly one solution $\mathbf{x} \in \mathbb{R}^L$. 
 \subsection{The solution expressed as eigenstate of an observable}  
 We first express the solution as eigenvector of an operator $A$ on a  L-dimensional Hilbert space $\mathcal{H}_L$ which is isomorphic to $\mathbb{C}^L$.  For this purpose, we use the transcription employed for the Variational Quantum Linear Solver (VQLS) algorithm \cite{Bravo-Prieto23}, i.e. 
 \begin{equation}
A \ket{y} = 0\cdot\ket{y} \equiv 0\,\,\mbox{with}\,\, A\equiv \hat{M}^\dagger \left(\mathbb{I} - \pro{b}\right)\hat{M}\,,
\label{eqnA}
\end{equation}
where  $\ket{y}=\sum_i y_i \ket{i}, \ket{b} = \sum_i b_i \ket{i} \in \mathcal{H}_L $ are normalised with $b_i$ the $i$-th component of vector $\mathbf{b}/\lVert \mathbf{b} \rVert$ and $\hat{M} = \sum_{ij} m_{ij} \ket{i}\bra{j}$ the operator on $\mathcal{H}_L$ corresponding to the matrix $M$ with elements $m_{ij}$ from  (\ref{eqnM}).  
If $\ket{y}$ satisfies eigenvalue equation (\ref{eqnA}), the solution of the linear system of equations (\ref{eqnM}) reads 
\begin{align} 
{\bf{x}} &= z (y_1, \ldots y_L)^T\,\, \mbox{with}\,\,  z = \frac{\lVert \mathbf{b} \rVert}{\bra{b}\hat{M}\ket{y} }\,. 
\label{solutionx}
\end{align}
This follows, since (\ref{eqnA}) is equivalent to 
 \begin{align}
\left(\mathbb{I} - \pro{b}\right)\hat{M}\ket{y} = 0 \Leftrightarrow \hat{M}\ket{y} = \ket{b} \bra{b}\hat{M}\ket{y}\,. 
\end{align}

The VQLS algorithm uses a function related to  $C(\mathbf{\psi})\equiv\bra{\psi}A\ket{\psi}/\bra{\psi}\hat{M}^\dagger \hat{M} \ket{\psi}$ as cost function to find $\ket{\psi}=\ket{y}$ by minimising the costs $C(\psi)$.  
In the following we derive a different objective function, and a new algorithm, {\sl the measurement test algorithm}, based on the fact that operator $A$ corresponds to a physical observable and can be measured, since it is self-adjoint, i.e.\ $A^\dagger = A$. Moreover, the solution of the linear system (\ref{eqnM}) corresponds to an eigenstate $\ket{y}$ of the observable $A$ with eigenvalue zero, cf.\ Eqn.\ (\ref{eqnA}).  
\subsection{Computation of the solution by measurement}  
The new algorithm is based on a quantum measurement (Von Neumann measurement) of the objective observable $A$. According to the theory of quantum measurement \cite{Busch.et.al91}, this involves a time evolution that couples the eigenstates $\ket{a_i} $ of the observable $A$ to a basis of pointer states  $\ket{\lambda_i}$ in one-to-one correspondence with the eigenvalues of the observable A, defined by $A\ket{a_i}=\lambda_i\ket{a_i}$,
\begin{equation}
\ket{a_i}\ket{0}\rightarrow \ket{a_i}\ket{\lambda_i} \,.
\end{equation}   
Given that  the initial state is prepared by a unitary $V$, i.e., $\ket{\psi(\alpha)}= V(\alpha)\ket{0}=\sum c_i(\alpha) \ket{a_i}$ with complex coefficients $ c_i(\alpha)$,  which depends on a set of parameters $\alpha\equiv\{\alpha_1, \alpha_2\ldots \alpha_n\}$, the state after the measurement coupling (the so-called {\it premeasurement}) reads
\begin{equation}
\ket{\Psi(\alpha)}= \sum c_i(\alpha) \ket{a_i}\ket{0}\rightarrow \sum c_i(\alpha)\ket{a_i}\ket{\lambda_i}\,.
\label{premeasurement}
\end{equation}  
The detection of the pointer value projects the apparatus into one of the pointer states $\ket{\lambda_i}$ (corresponding to the  {\it objectification} part of the measurement), and thus prepares the corresponding eigenstate of the observable,
\begin{align}
\sum c_i(\alpha)\ket{a_i}\ket{\lambda_i}  \xrightarrow{\lambda_i} \ket{a_i}\ket{\lambda_i}\,.
\label{objectification}
\end{align}
In this way any eigenstate of the observable $A$ can be prepared  by the quantum algorithm, provided it implements a measurement of $A$ with the corresponding measurement result (pointer value). The readout of the pointer values occurs    
with probability 
\begin{equation}
p(\lambda_i)=\langle \mathbb{I} \otimes  \ket{\lambda_i}\bra{\lambda_i}\rangle_\Psi = \langle \ket{a_i}\bra{a_i}\rangle_\psi =
\vert c_i(\alpha)\vert^2. 
\label{bla}
\end{equation}  
For the minimal pointer value $\lambda_0=0$ this is equal to the target fidelity $F_T$, i.e.,
\begin{equation}
p(0)=\vert\bra{y}\psi(\alpha)\rangle\vert^2\equiv F_T\,,
\label{blabla}
\end{equation}
and the desired information, the solution $\mathbf{x}$, is encoded in the target state $\ket{y}\equiv\ket{a_0} $. Hence, this information can be directly obtained from the optimal values $\alpha=\alpha_{\mathrm{opt}}$ by optimising the detection probability $p(0)\le 1$ to detect $\lambda_0=0$.  This follows from
\begin{equation}
\ket{\psi(\alpha)}=\ket{y}\,\, \Leftrightarrow \,\, p(0)= 1\,.
\label{ketx}
\end{equation}
Therefore, the solution of the linear system (\ref{eqnM}), given by the eigenstate $\ket{y}$  of the observable  $A$, can be computed by directly maximising the detection probability $p(0)$, which represents the merit function of the measurement test algorithm.

  \subsection{Ansatz} 
  \label{ansatz}
  The set of unitary gates $V(\alpha)$, obtained by varying the parameters $\alpha$, 
  that prepare the initial state 
 \begin{equation}
  \ket{\psi(\alpha)}= V(\alpha)\ket{0}
  \end{equation}
in the input register, needs to be sufficiently large to generate the target state $\ket{y}= V(\alpha)\ket{0}$  for a particular $\alpha= \alpha_{\mathrm{opt}}$. The parameters of the so-called Ansatz $\ket{\psi(\alpha)}$ are  optimised in the measurement test algorithm until  the convergence condition (\ref{ketx}) is met with a certain accuracy $\epsilon$, i.e., the relative frequency of the measurement results $\lambda_0=0$ indicate $p(0)=1-\epsilon$.  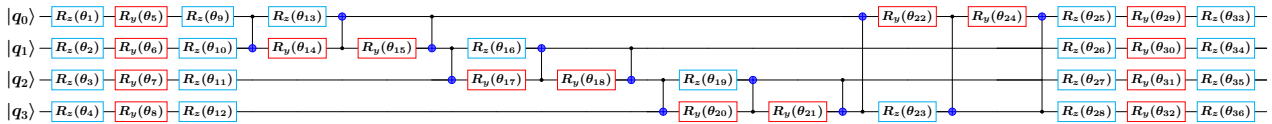
\begin{figure*}
\tikzset{
	mycnot/.style={fill=blue!60, draw=blue}
}

\scalebox{0.32}{
	\begin{quantikz}
		\lstick{\huge\(\bm{|q_0\rangle}\)} & \gate[style={fill=cyan!0, draw=cyan!90!cyan}]{\scalebox{1.7}{$\bm{R_z(\theta_1)}$}} & \gate[style={fill=red!0, draw=red!90!red}]{\scalebox{1.7}{$\bm{R_y(\theta_5)$}}}& \gate[style={fill=cyan!0, draw=cyan!90!cyan}]{\scalebox{1.7}{$\bm{R_z(\theta_9)}$}} & \ctrl[style=fill=mycnot]{1} &  \gate[style={fill=cyan!0, draw=cyan!90!cyan}]{\scalebox{1.7}{$\bm{R_z(\theta_{13})}$}}\qw&
		\targ[style=mycnot]& & &
		\ctrl[style=fill=mycnot]{1}& && &&&&&&&&& 
		\targ[style=mycnot]&\qw&
		\gate[style={fill=red!0, draw=red!90!red}]{\scalebox{1.7}{$\bm{R_y(\theta_{22})}$}} &
		\ctrl[style=fill=mycnot]{3}&
		\gate[style={fill=red!0, draw=red!90!red}]{\scalebox{1.7}{$\bm{R_y(\theta_{24})}$}}&
		\targ[style=mycnot]&\qw&
		\gate[style={fill=cyan!0, draw=cyan!90!cyan}]{\scalebox{1.7}{$\bm{R_z(\theta_{25})}$}} & \gate[style={fill=red!0, draw=red!90!red}]{\scalebox{1.7}{$\bm{R_y(\theta_{29})}$}}& \gate[style={fill=cyan!0, draw=cyan!90!cyan}]{\scalebox{1.7}{$\bm{R_z(\theta_{33})}$}} &
		\qw \\
		\lstick{\huge\(\bm{|q_1\rangle}\)} & \gate[style={fill=cyan!0, draw=cyan!90!cyan}]{\scalebox{1.7}{$\bm{R_z(\theta_2)}$}} & \gate[style={fill=red!0, draw=red!90!red}]{\scalebox{1.7}{$\bm{R_y(\theta_6)}$}} & \gate[style={fill=cyan!0, draw=cyan!90!cyan}]{\scalebox{1.7}{$\bm{R_z(\theta_{10})}$}} & \targ[style=mycnot] & \qw & \gate[style={fill=red!0, draw=red!90!red}]{\scalebox{1.7}{$\bm{R_y(\theta_{14})}$}}& \ctrl[style=fill=mycnot]{-1} &
		\gate[style={fill=red!0, draw=red!90!red}]{\scalebox{1.7}{$\bm{R_y(\theta_{15})}$}}&
		\targ[style=mycnot] &\qw&
		\ctrl[style=fill=mycnot]{1} &
		\gate[style={fill=cyan!0, draw=cyan!90!cyan}]{\scalebox{1.7}{$\bm{R_z(\theta_{16})}$}} &
		\targ[style=mycnot] &\qw&&
		\ctrl[style=fill=mycnot]{1} &&&&&&&&&&&\qw&
		\gate[style={fill=cyan!0, draw=cyan!90!cyan}]{\scalebox{1.7}{$\bm{R_z(\theta_{26})}$}} & \gate[style={fill=red!0, draw=red!90!red}]{\scalebox{1.7}{$\bm{R_y(\theta_{30})}$}}& \gate[style={fill=cyan!0, draw=cyan!90!cyan}]{\scalebox{1.7}{$\bm{R_z(\theta_{34})}$}} &
		\qw\\
		\lstick{\huge\(\bm{|q_2\rangle}\)} & \gate[style={fill=cyan!0, draw=cyan!90!cyan}]{\scalebox{1.7}{$\bm{R_z(\theta_3)}$}} & \gate[style={fill=red!0, draw=red!90!red}]{\scalebox{1.7}{$\bm{R_y(\theta_{7})}$}} & \gate[style={fill=cyan!0, draw=cyan!90!cyan}]{\scalebox{1.7}{$\bm{R_z(\theta_{11})}$}}  & &&&&&
		\targ[style=mycnot] &\qw& \gate[style={fill=red!0, draw=red!90!red}]{\scalebox{1.7}{$\bm{R_y(\theta_{17})}$}}&
		\ctrl[style=fill=mycnot]{-1} &
		\gate[style={fill=red!0, draw=red!90!red}]{\scalebox{1.7}{$\bm{R_y(\theta_{18})}$}}&
		\targ[style=mycnot] &\qw&&
		\ctrl[style=fill=mycnot]{1} &
		\gate[style={fill=cyan!0, draw=cyan!90!cyan}]{\scalebox{1.7}{$\bm{R_z(\theta_{19})}$}} &
		\targ[style=mycnot] &\qw&&
		\ctrl[style=fill=mycnot]{1} & &&&&\qw&
		\gate[style={fill=cyan!0, draw=cyan!90!cyan}]{\scalebox{1.7}{$\bm{R_z(\theta_{27})}$}} & \gate[style={fill=red!0, draw=red!90!red}]{\scalebox{1.7}{$\bm{R_y(\theta_{31})}$}}& \gate[style={fill=cyan!0, draw=cyan!90!cyan}]{\scalebox{1.7}{$\bm{R_z(\theta_{35})}$}} &
		\qw\\
		\lstick{\huge\(\bm{|q_3\rangle}\)} & \gate[style={fill=cyan!0, draw=cyan!90!cyan}]{\scalebox{1.7}{$\bm{R_z(\theta_4)}$}} & \gate[style={fill=red!0, draw=red!90!red}]{\scalebox{1.7}{$\bm{R_y(\theta_8)}$}} & \gate[style={fill=cyan!0, draw=cyan!90!cyan}]{\scalebox{1.7}{$\bm{R_z(\theta_{12})}$}} & &&&&& &&&&&& 
		\targ[style=mycnot] &\qw&
		\gate[style={fill=red!0, draw=red!90!red}]{\scalebox{1.7}{$\bm{R_y(\theta_{20})}$}} &
		\ctrl[style=fill=mycnot]{-1} &
		\gate[style={fill=red!0, draw=red!90!red}]{\scalebox{1.7}{$\bm{R_y(\theta_{21})}$}} &
		\targ[style=mycnot]&\qw &
		\ctrl[style=fill=mycnot]{-3}&
		\gate[style={fill=cyan!0, draw=cyan!90!cyan}]{\scalebox{1.7}{$\bm{R_z(\theta_{23})}$}} &
		\targ[style=mycnot]&\qw & &
		\ctrl[style=fill=mycnot]{-3}&
		\gate[style={fill=cyan!0, draw=cyan!90!cyan}]{\scalebox{1.7}{$\bm{R_z(\theta_{28})}$}} & \gate[style={fill=red!0, draw=red!90!red}]{\scalebox{1.7}{$\bm{R_y(\theta_{32})}$}}& \gate[style={fill=cyan!0, draw=cyan!90!cyan}]{\scalebox{1.7}{$\bm{R_z(\theta_{36})}$}} &
	\end{quantikz}
}
	\caption{Ansatz with $36k$ rotation angles generated by $k$ sequential applications of  the module depicted.     }
	\label{fig1}
\end{figure*}

In order to demonstrate the capacity of the measurement test algorithm, we solve an arbitrary linear system of 16 equations in 16 variables. For this purpose, we choose the shallow ansatz module displayed in Fig.~\ref{fig1}. This module is inspired by approximate unitary designs \cite{Schuster.et.al25} and can be iterated to increase the number of angles for expressibility. It consists of two layers of general single qubit rotations with three variable Euler angles sandwiching $C$-$X$ gates between next neighbours in a circular configuration with additional rotations.  
 
 \subsection{Implementation of "textbook" quantum measurement} 
 \label{PEA}
\begin{figure}
\tikzset{dottedwire/.style={dash pattern=on 1pt off 2pt}}
\scalebox{0.44}{
	\begin{quantikz}
		\lstick{\Large\(\bm{|a_0\rangle}\)} & 
		\gate[style={fill=cyan!0, 
			draw=cyan!90!cyan}]{\scalebox{1.7}{$H$}}& &&
		\ctrl[style=fill=mycnot]{4}& &&&&&&&&&&&&
		\gate[wires=4, style={fill=red!20, draw=red!80!black}]{\scalebox{1.7}{$QFT^{\dagger}$}}&&&
		\meter{} &
		\qw\\ 
		\lstick{\Large\(\bm{|a_1\rangle}\)} & \gate[style={fill=cyan!0, draw=cyan!90!cyan}]{\scalebox{1.7}{$H$}}& &&&&&&&
		\ctrl[style=fill=mycnot]{3}& &&&&&&&&&&
		\meter{} &
		\qw\\
		\lstick{}  \push{\makebox[1cm][l]{\bm{\vdots} \hspace{11cm} }}   \\
		\lstick{\Large\(\bm{|a_{m-1}\rangle}\)} &
		\gate[style={fill=cyan!0, draw=cyan!90!cyan}]{\scalebox{1.7}{$H$}} & &&&&&&&&&&\push{\tikz{\,\,\cdots\,\,}}&&& 
		\ctrl[style=fill=mycnot]{1}& &&&&
		\meter{} &
		\qw\\ 
		\lstick{\Large\(\bm{|s_0\rangle}\)} & 
		\gate[wires=4, style={fill=blue!20,
			minimum width=2.5cm,
			minimum height=1.5cm, draw=blue!80!black}]{\scalebox{1.7}{$V(\theta)$}} & &&\gate[wires=4, style={fill=cyan!20, draw=green!80!black}]{\scalebox{1.7}{$U^{2^0}$}}& &&& &\gate[wires=4, style={fill=cyan!20, draw=green!80!black}]{\scalebox{1.7}{$U^{2^1}$}}&&&\push{\tikz{\,\,\cdots\,\,}}&&&\gate[wires=4, style={fill=cyan!20, draw=green!80!black}]{\scalebox{1.7}{$U^{2^{m-1}}$}}&&&&&& \\
		\lstick{\Large\(\bm{|s_1\rangle}\)} &&&&  \qw&&&&&&&&\push{\tikz{\,\,\cdots\,\,}}&&&&&&&&& \\
		\lstick{}  \push{\makebox[1cm][l]{\bm{\vdots} \hspace{11cm}}}   \\
		\lstick{\Large\(\bm{|s_{n-1}\rangle}\)} &&&& \qw&&&&&&&&\push{\tikz{\,\,\cdots\,\,}}&&&&&&&&& 
		\end{quantikz}
	}
	\caption{Quantum circuit of measurement test algorithm.  Ansatz $V(\alpha)$ in input register and  quantum gates of phase estimation algorithm, as well as measurements in output register. The relative frequency of "$0$" results serves as estimator of the probability $p_0$ to be maximised by classical optimisation. }\label{fig2}
	\end{figure}
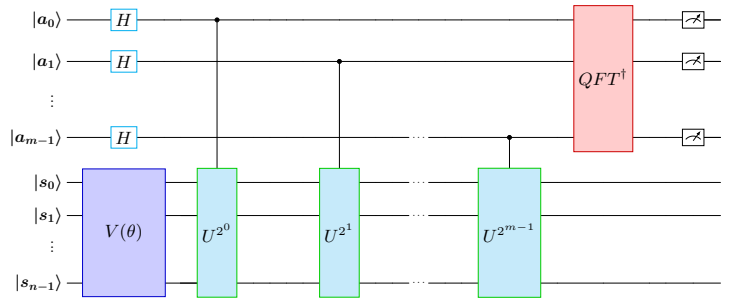

The measurement of the observable $A$ is realised by applying the phase estimation algorithm  \cite{Nielsen.Chuang00} to the ansatz state,  cp.\ Fig.\ \ref{fig2}, using an input and an output register of $n, m$ qubits, respectively. For this purpose, we assume below that the phase estimation algorithm can be implemented for $A$, which requires fault tolerant quantum computing and might pose additional constraints to $A$, see Section \ref{discussion}. The output register (upper register) is initialised by local Hadamard gates $H$ in a superposition of $K=2^m$ computational basis states, $\sum_k \ket{k}/\sqrt{K}$. Conditioned on the state of the output register, the input register is transformed by a power of the unitary $U=\exp(2\pi i A)$, where $A$ is the observable to be measured. 
 \begin{align}
 \frac{1}{\sqrt{K}}\sum_{k=0}^{K-1}\ket{a_i} \ket{k}\xrightarrow{}& \frac{1}{\sqrt{K}}\sum_{k=0}^{K-1}  (U^k\ket{a_i})\ket{k}\nonumber\\
& =\ket{a_i}  \frac{1}{\sqrt{K}}\sum_{k=0}^{K-1} e^{2\pi i \lambda_i k} \ket{k}
 \label{operationPE1}
 \end{align}
The eigenvalue $\lambda_i\in[0,1]$ to eigenvector $\ket{a_i}$ can be expressed as $\lambda_i=j/K +R$  with $j\in\{0,1,\dots, K-1\}$, where a remainder $R\le 1/2^{m}$ is the resolution limit for $m$ output qubits and is neglected below. For the successful detection of $\lambda_0=0$ this is justified if all other eigenvalues are greater than the resolution limit. This condition is related to a logarithmic scaling of the number of qubits with the condition number of matrix $M$, as shown in Section \ref{section3}.

An inverse Quantum Fourier Transform acting on the output register $\ket{k} \rightarrow \sum_l e^{-\frac{2\pi i k l}{K}} \ket{l}/\sqrt{K}$  leads to 
 \begin{align}
  &\ket{a_i}  \frac{1}{\sqrt{K}}\sum_ke^{\frac{2\pi i j k}{K}} \ket{k} \nonumber\\
 & \xrightarrow{\mathrm{QFT}^{-1}}  \frac{1}{K}\ket{a_i} \sum_{k,l}e^{\frac{2\pi i (j-l) k}{K}}\ket{l} = \ket{a_i}\ket{\lambda_i},
 \label{operationPE2}
 \end{align}
 where $\ket{\lambda_i}\equiv\ket{j} = \ket{b_{0} b_{1}\ldots b_{m-1}} $ refers to the binary fractional representation of the eigenvalue, i.e.,  
 \begin{equation}
  \lambda_i= \frac{j}{K} = {b_0}2^{-1}+ {b_1}2^{-2} \ldots b_{m-1}2^{-m}
 \label{lambdabits}
 \end{equation}
 with binary digits $b_n\in\{0,1\}$. Linear operations (\ref{operationPE1})  and (\ref{operationPE2}), when applied to an arbitrary state, thus implement the pre-measurement (\ref{premeasurement}), while a subsequent measurement of the qubits of the output register in the computational basis results in bit values $(b_{0},\ldots b_{m-1})$ that identify directly an eigenvalue of the observable via binary expansion (\ref{lambdabits}) within the resolution limit, $R=1/2^{m}$, imposed by the number $m$ of qubits of the output register. 
 
In terms of the measured system the measurement result (eigenvalue of observable), final state (eigenvector) and probability reflect the measurement postulate found in textbooks on quantum mechanics
  \begin{equation} 
 \ket{\psi}\xrightarrow{\lambda_i} \ket{a_i}\,\,\mbox{with probability}\,\, p(\lambda_i)= |\bra{a_i} \psi \rangle\vert^2\,.
 \end{equation}
We note that the pointer values in other measurement models, e.g. the standard model of measurement \cite{Busch.et.al91}, depend on the coupling strength and are usually not equal to the eigenvalues of the observable. It is therefore remarkable that this way to implement a measurement of an observable yields exactly its eigenvalues as results.  
 
 In the measurement test algorithm, the measurements serve to estimate the probability to detect the eigenvalue $\lambda_0 = 0$. For this purpose,  the quantum circuit displayed in Fig.~\ref{fig2} is run $N$ times and measurements of the $m$ qubits in the output register are carried out, recording the number $N_0$ of times that all qubits show a zero result, which represents the zero eigenvalue  $\lambda_0 =(0 0 \cdots 0)$ in binary digits (\ref{lambdabits}). The relative frequency $N_0/N$ serves as estimator  of $p_0$. Its properties are studied in Section \ref{section3}.
  
 \subsection{Optimisation and Convergence to Target} 
The estimated probabililty is optimised using a classical optimisation algorithm by modifying the parameters of the input state to obtain $p(0)=1$ and thus generate the parametrisation of the solution $\ket{y}$ due to the equivalence (\ref{ketx}). For this purpose, we chose the optimisation algorithm Rotosolve \cite{rotosolve} that we found works best for the measurement test algorithm. After each application of the optimisation algorithm a new iteration of the measurement test algorithm is executed until the solution is approximated with  a certain accuracy $\epsilon$. 

\begin{figure}
	\includegraphics[width=0.5\textwidth]{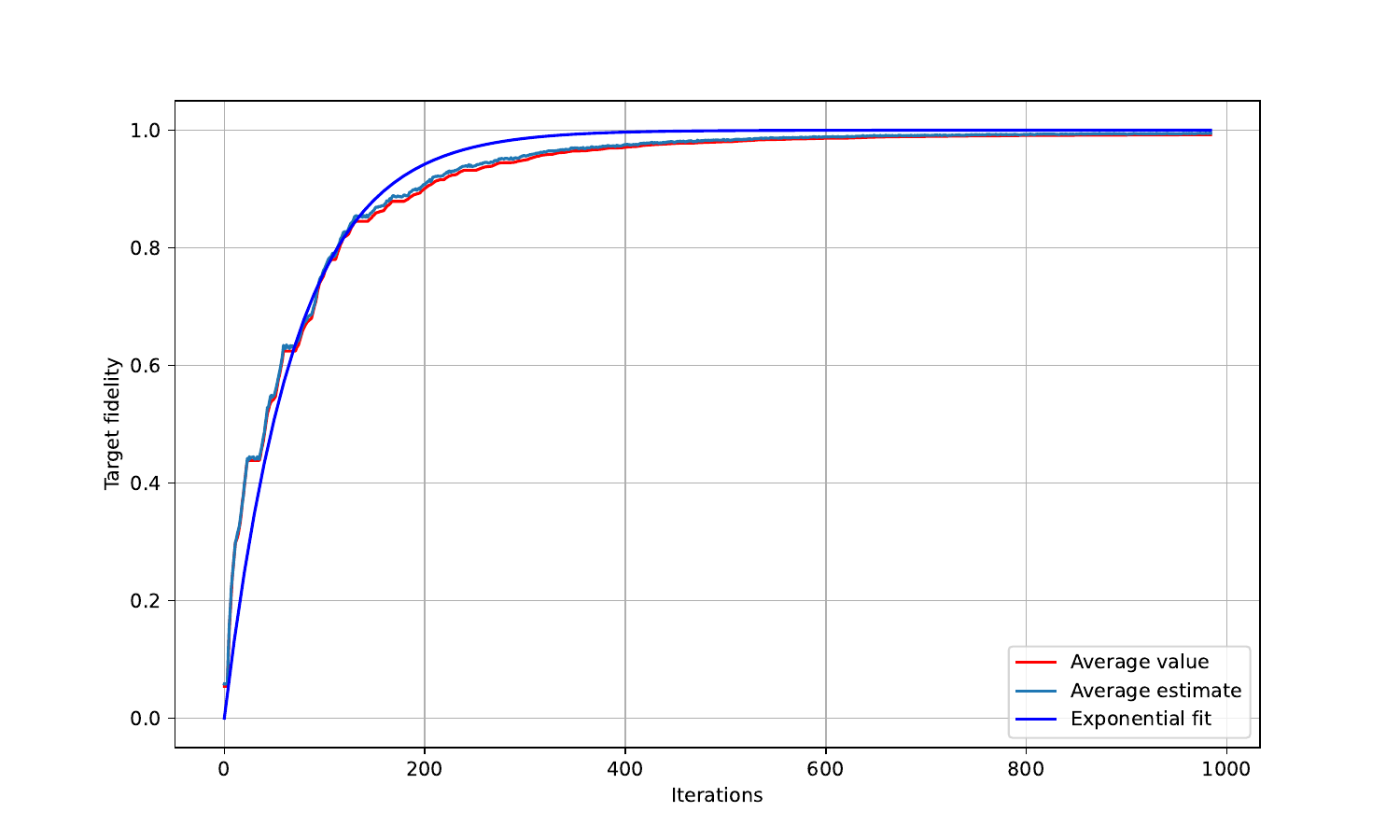}
	\caption{Numerical simulation show convergence of target fidelity (red) and its estimate (green) averaged over fifty runs for different random matrices $M$ and $N=1000$ measurements per iteration $t$. Best fit $f(t) = 1- e^{-\gamma t})$ (blue)  with parameter $\gamma= 0.014$  demonstrates  exponentially fast rise and then saturates at unity. The asymptotic behaviour of $F_T$ is analysed in Sect.~\ref{section3}).}
\label{fig3a}
\end{figure}

\begin{figure}
	\includegraphics[width=8cm]{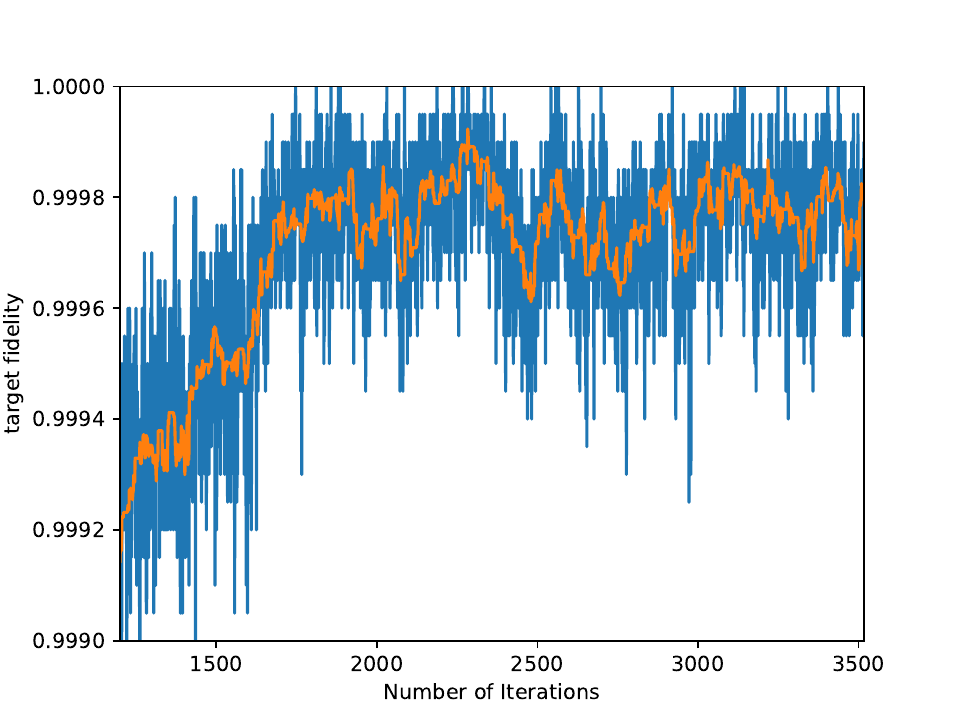}
	\caption{ Target fidelity for simulation of a single run with $N=20000$ shots per iteration. The measured  relative frequency (blue) saturates when it sufficiently often reaches the maximum value of $1$, while the actual target fidelity (red) is still less than $1$. } 
	\label{fig3b}
\end{figure}

\begin{figure}
	\includegraphics[width=8cm]{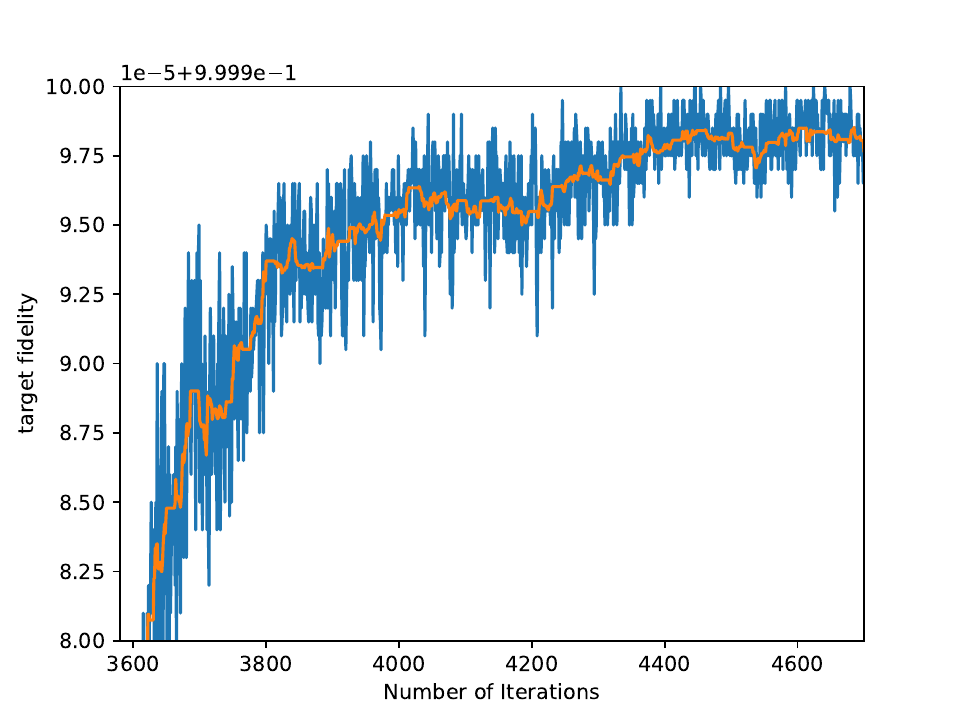}
	\caption{Subsequent section to the one shown in Fig.\ref{fig3b} with $100$ times as many measurements per iteration. The accuracy $\epsilon$ ($F_T=1-\epsilon$) is two orders of magnitude smaller than in Fig.~\ref{fig3b}. Similar to there, the target fidelity  increases exponentially until the relative frequencies reach unity often. }  
	\label{fig3c}
\end{figure}

Figures \ref{fig3a}-\ref{fig3c} show the characteristic features of convergence towards the target state $\ket{y}$ obtained from numerical simulations of the measurement test algorithm for different random dense  $16\times16$ matrices $M$, that are invertible and symmetric, as well as its corresponding observables $A$. In order to guarantee expressibility, the ansatz of the measurement test algorithm was here composed of three iterations of the module in Fig.~\ref{fig1}.  For each matrix the target fidelity (Fig.~\ref{fig3a}) increases at first exponentially with the number of iterations of the measurement test algorithm until it  saturates when the detected relative frequencies reach unity and the solution  $\mathbf{x}$ of the linear system of equations (\ref{eqnM}) can be reconstructed with a certain accuracy.   
The maximal target fidelity the measurement test algorithm yields for sufficiently many iterations is given by 
\begin{equation} 
F_T\equiv p(0) = 1 - \epsilon\,\,\mbox{with}\,\, \epsilon \propto \frac{1}{N}\,,  
\label{maxfidel}
\end{equation} 
where $N$ is the number of repeated measurements, cf.\ Section \ref{section3}. This Heisenberg scaling \cite{Hayashi.et.al22} of the accuracy $\epsilon$ is observed in the numerical simulation shown in 
Fig.~\ref{fig3b} where the target fidelity (red)  is levelling off at $F_T= 1- 2\times10^{-4}$ for $N=2\times10^4$ shots after 2000 iterations. The fluctuating estimates of $p(0)$ (blue) frequently display maximal (unit) values,  for which the classical optimiser no longer modifies the state parameters. This might lead to the saturation of target fidelity.  

In order to test this hypothesis, after the target fidelity had not improved during the last 1500 iterations shown in Fig.~\ref{fig3b}, the number of shots was subsequently increased to $N=2 \times 10^6$. As displayed in Fig.~\ref{fig3c}, the increase resulted in the convergence to the new stationary value of $F_T= 1- 2\times 10^{-6}$ within additional $800$ iterations. This increase of target fidelity indeed indicates that here the number of measurements determines the accuracy. 
 \section{Accuracy of the Solution }
 \label{section3}
 There are three necessary conditions for convergence to the solution through iterations of the measurement test algorithm.  (a) The ansatz must express the solution, i.e.  $\ket{\psi(\alpha)}= \ket{y}$ for specific $\alpha$.   (b) the measurement must discriminate the second smallest eigenvalue $\lambda_1$ of the observable $A$ from its smallest one $\lambda_0=0$, and (c) the optimiser must be able to find the solution with a required accuracy within a reasonable number of steps.  While the first and last condition, (a) and (c), are based on heuristics, they are not viewed as critical here. Our choices for both - ansatz and optimiser - found the solution up to a certain accuracy for all  random dense $16\times16$ matrices that we studied (more than 100) within $10^4$ iterations. To improve ansatz or optimiser is left for further exploration elsewhere.
The second condition (b)  can be satisfied with logarithmic resources in terms of the required number of qubits for the pointer states allocated in the output register.    
 
For a random square matrix $M$ with condition number $\kappa$, related to observable $A$ by Eq.\ (\ref{eqnA}), the second smallest eigenvalue of $A$ is on average  of the order of magnitude $\lambda_1\approx 1/\kappa^2$ \cite{Bravo-Prieto23}. Since the resolution $R=1/2^{m}$ of the measurement of eigenvalues is determined by the number $m$ of qubits in the output register, see Section \ref{PEA},  we can relate $m$ to the condition number of matrix $M$
\begin{equation}
\lambda_1\ge R\,\, \Leftrightarrow\,\, \frac{1}{\kappa^2}\ge \frac{1}{2^{m}}\,\, \Leftrightarrow\,\,  m\ge 2\log_2(\kappa).
\end{equation}

Assuming conditions (a)-(c) are satisfied, the accuracy of the fault-tolerant algorithm (or its simulation) is still constrained by the shot noise inherent in measuring the target fidelity to be optimised. 
      
For the measurement of observable $A$, we only distinguish between projecting on eigenstate $\ket{y}$ (result  $\lambda_0 =0$) and projecting on its orthogonal complement (result $\lambda \not=0$).  In passing we note that this is equivalent to measuring the observable $P_0= \ket{y}\bra{y}$ on the system.  
The probability $p_0 \equiv p(0)$  to measure $0$ is given by  Eqn.\ (\ref{blabla}), and the counter probability for $\lambda \not=0$ thus reads $(1-p_0)$. 

\begin{figure}
	\includegraphics[width=8cm]{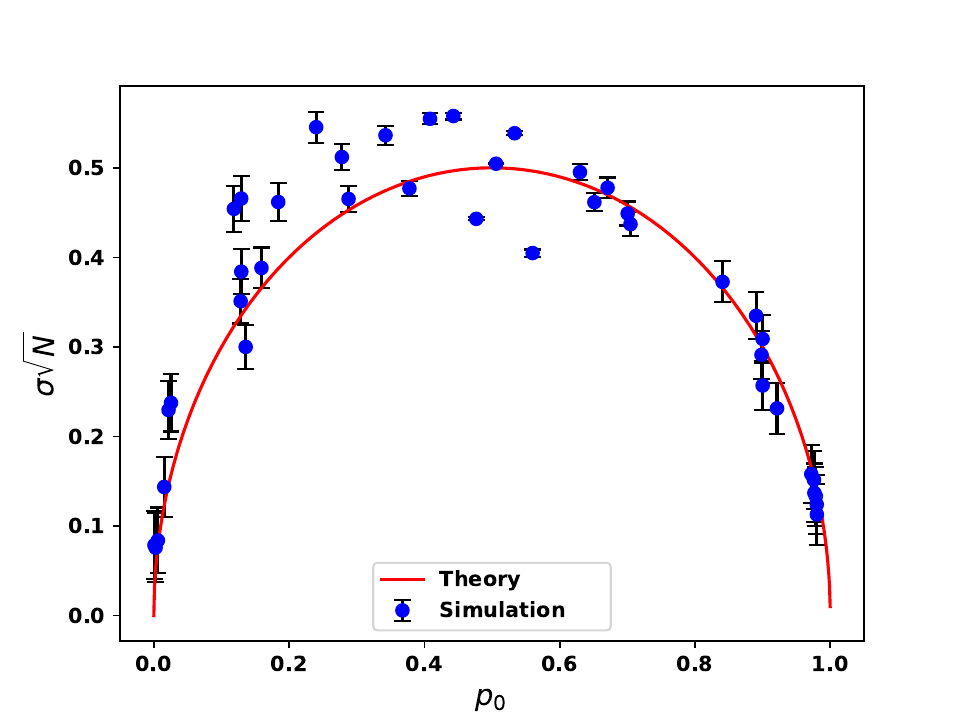}
	\caption{ Standard deviation of estimator for various values of probability $p_0$ as detected in numerical simulations of the measurements in the algorithm.} 
	\label{fig5}
\end{figure}

Without restriction of generality, let us assume that the result $\lambda_0$ occurs $N_0$ times in  $N$ measurements.  Then the relative frequency $r=N_0/N$ follows  a binomial probability distribution, 
\begin{align}
p(r=N_0/N) &=  \binom{N}{ N_0} p_0^{N_0}(1-p_0)^{N-N_0} \nonumber \\ 
&\approx \frac{1}{\sqrt{2\pi} N\sigma(r)} e^{-\frac{(r-p_0)^2}{2\sigma^2(r)}},
\end{align}
and serves as an optimal (so-called efficient) estimator of $p_0$, since there is no other estimator with expectation value equal to the estimated quantity, $p_0= {\bar r} $, and a smaller variance \cite{Kreyszig65} than 
\begin{equation}
\sigma^2(r)= \frac{p_0(1-p_0)}{N}.
\end{equation}
We tested our implementation  of the quantum measurement (cf.~Section \ref{PEA}) in simulations of the quantum circuit that reproduced the dependence of the variance on  the probability to detect $\lambda_0 =0$ correctly, see Fig.\ \ref{fig5}.   

With increasing multiples $a$ of the standard deviation $\sigma$, a larger proportion of the unbiased estimator's $r=N_0/N$ distribution is contained within an interval given by 
\begin{equation}
r = p_0\pm a \sigma\le 1  \label{approx}
\end{equation}
If the conditions (a)-(c) for convergence are met, the relative frequency increases over multiple iterations of the measurement test algorithm until it reaches its maximum value, i.e.~$r=1$, where the parameters $\alpha$ of the ansatz state $\ket{\psi(\alpha)}$ and thus the value of merit function $p_0=\vert\bra{a_0}\psi(\alpha)\rangle\vert^2$ are not further optimised.  We assume here that  $p_0$ reaches an assymptotic value  (modulo fluctuations) if, and only if, this happens sufficiently often. The maximum value of $p_0 \le1$  that the measurement test algorithm thus yields reads       
\begin{align}
 & p_0 = 1- a\sigma  = 1- a\sqrt{\frac{p_0(1-p_0)}{N}}\,\,   \Leftrightarrow \nonumber\\
 & a^2 p_0 = N(1-p_0) \,\,  \Leftrightarrow\,\, p_0 =\frac{N}{N+a^2} \approx 1 - \frac{a^2}{N}.    \label{p0b}
\end{align}
The approximation in the last line of (\ref{p0b}) is valid if $a^2\ll N$. This is related to the condition for termination of the optimisation, since the probability to detect $r\approx 1$ is the greater, the smaller the value of $a$. 

If $a$ is constant, independent of the number $N$ of shots,  then the accuracy of the target fidelity $F_T\equiv p_0 =1-\epsilon$ shows Heisenberg scaling $\epsilon \approx a^2/N$, cf.~(\ref{p0b}). Both is confirmed by the assymptotic values of target fideliy $F_T\equiv p_0 $ that we obtained from numerical simulations of the presented algorithm, cf.\ Table \ref{table2}.
\begin{table}
	\centering
	\caption{ Scaling of asymptotic target fidelity $F_T$ with number of measurements (shots) $N$.\label{table2}}
	\begin{ruledtabular}
		\begin{tabular}{cccccc}
			$N$ shots	&100&1000&$10^4$&$10^5$& $10^6$\\
			$F_T$	& 0.953 & 0.9948 & 0.99942 & 0.99995& 0.9999946\\
			$1 - 4/N$ & 0.96 &  0.996 & 0.9996 & 0.99996& 0.999996\\	
			Value of $a$ & 2.17 & 2.28 & 2.41 & 2.23 & 2.32 \\	
		\end{tabular}
	\end{ruledtabular}
\end{table}
While similar values of $2\le a\le 2.5$ were obtained for different random dense matrices $M$, $a$ might depend on the optimiser used.   


We  note that with fault tolerant quantum computing the measurement test algorithm permits the solution of a linear system of equations with a dense instead of a sparse matrix $M$, because the size of the standard deviation $\sigma$ of the estimator does not - as in other methods (for example VQLS) - depend on the number of Pauli strings that compose the matrix $M$. 
This dependence arises when the observable of the merit function is expressed in terms of Pauli strings, the expectation value of each Pauli string is measured individually and accumulates additional shot noise.  
    
\begin{figure}
	\includegraphics[width=8cm]{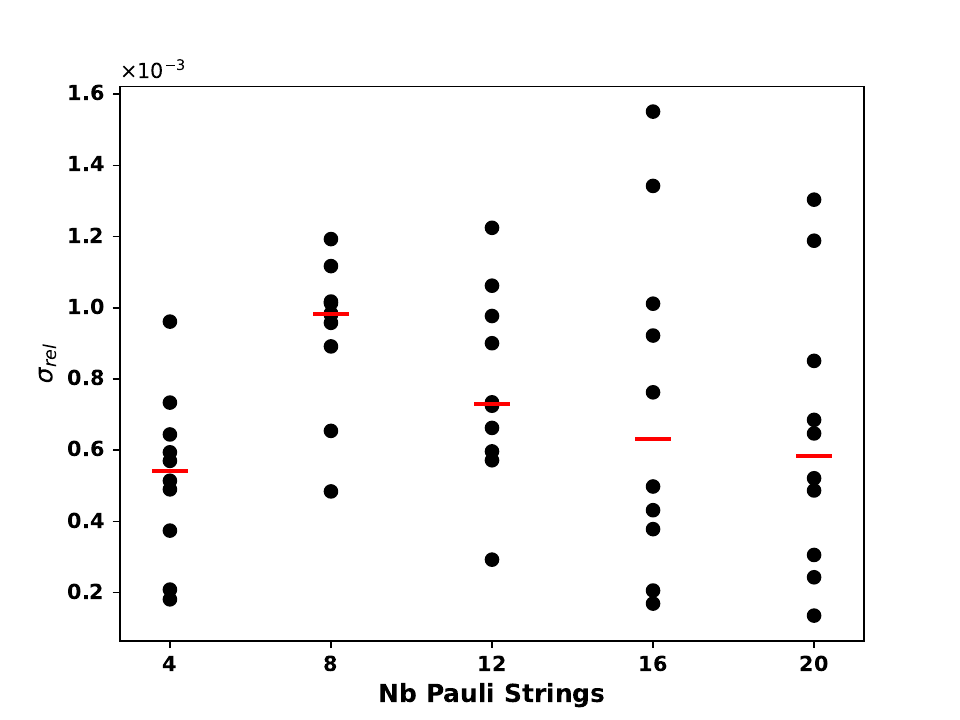}
	\caption{ The median relative standard deviation  $\sigma_{\rm rel}$ (red line) for a total number of $N= 100 000$ measurements, cf.~ main text, does not increase with the number of Pauli strings of $M$ . 
	}
	\label{fig7}
\end{figure}

\begin{figure}
	\includegraphics[width=8cm]{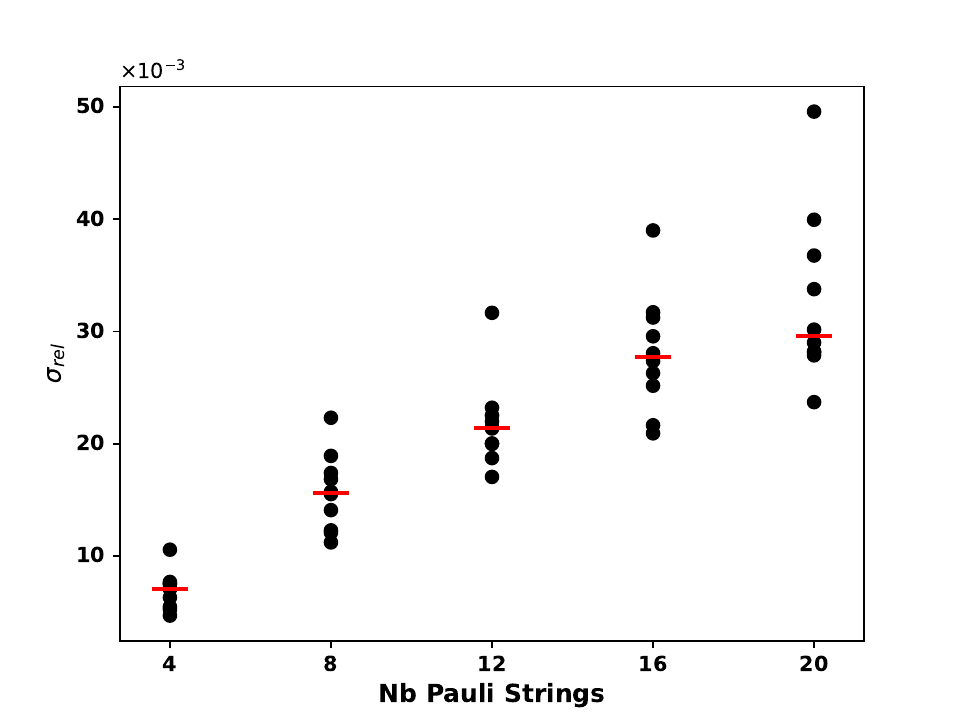}
	\caption{
The median relative standard deviation  $\sigma$ (red line), cf.\ main text,  of the cost function $C$  of VQLS increases with the number of Pauli terms of $M$}. 
	\label{fig8}
\end{figure}

Fig.\ \ref{fig7} and Fig.\ \ref{fig8} compare for various numbers of Pauli strings of matrix $M$ the resulting standard deviation of the estimator of  $p_0$  with the standard deviation of the corresponding expression $C(\mathbf{\psi})$ in VQLS (cf.~Section \ref{section2}), respectively. The results from numerical simulations of the quantum measurements for ten random matrices $M$ for each number of Pauli strings with fifty repetitions each, show smaller non-increasing values of the relative standard deviation $\sigma_{\rm rel}\equiv\sigma/p_0$ of the measurement test algorithm, while the corresponding relative standard deviation $\sigma_{\rm rel}\equiv \sigma/C(\psi)$ of VQLS grows approximately linear with the number of Pauli strings. 
For a generic  $16\times 16$ matrix one would need up to 256 Pauli strings, far too many to detect $C(\mathbf{y})$ accurately on the computers used for this analysis in a reasonable time. In general, in order to obtain a comparable accuracy as with the measurement test algorithm, it would take $4^n \times 4^n$ times more measurements using the Hadamard test, since the evaluation of the cost function $C$ features this many Pauli basis strings for a dense $M$ encoded with $n$ qubits. This scaling excludes VQLS for high-dimensional applications and indicates why the capability to deal with dense matrices means a great step ahead.   


 \section{Discussion}
 \label{discussion}
 We  designed the measurement test algorithm to solve a linear system of equations by a quantum measurement that projects onto an eigenstate  which carries the solution in amplitude encoding.  While the target state can be prepared in this way, its read out is facilitated by varying the parameters of the input state using a classical optimiser to reach maximum probability to project onto the target state.  For unit probability the initial state equals the target state and the solution of the linear system of equations can be inferred from the optimal parameters.   In general, computing solutions in amplitude encoding, which enables storage capacity that grows exponentially with the number of qubits,  requires such mechanisms to read out the amplitudes efficiently   -- also on fault tolerant quantum computers.

 We emphasise the difference to other variational quantum algorithms, which employ a quantum computer to measure the expectation value of a known observable and optimise it. In contrast, our algorithm measures and optimises the expectation value of an unknown observable (the projector onto the target state). 
This constitutes a quantum algorithm without classical analog. Since a Turing machine can only evaluate known functions with an argument given as input, the evaluation of a function unknown to the programmer might constitute a quantum advantage beyond exponential speed ups, and pave the way to new applications. 

We demonstrated at the example of a system of 16 equations and variables with a dense matrix $M$ with random elements and unique solution, that for fault tolerant quantum computing the accuracy of the solution computed by the measurement test algorithm is only limited by the shot noise of the measurement of the eigenvalues, given that storage size allows to distinguish the smallest from the second smallest eigenvalue. For a matrix $M$ with condition number $\kappa$ this is the case for a register with a number $m>2\log\kappa$ of qubits.  Our results indicate that 
the target fidelity $F_T$ reaches exponentially fast (in terms of iterations) values closed to $F_T = 1-4/N$, where $N$ is the number of repetitions of the measurement per iteration.   In terms of applicability an important question is, how the number of iterations  to reach the asymptotic value of target fidelity changes depending on the dimensions of $M$. First tests for random linear systems with $1024$ equations and variables show that on computers with $200-300$ Gigaflops it takes a few days to reach a target fidelity of $20\%$ as opposed to a few hours to yield $F= 1-10^{-5}$ for $16$ equations. 

In contrast to classical algorithms to solve linear systems, the accuracy of the measurement algorithm, $\epsilon \approx 4/N$,  does not depend on the condition number. Provided that a solution is found to facilitate the search of vast parameter spaces this could lead to an advantage for solving linear systems of equations by means of simulations of the measurement test algorithm on a classical computer.   

We assumed above that the phase estimation algorithm  can be implemented for observable $A$, which implies $A$ to be such that the unitary $\exp( 2\pi i A)$ can be realised on a quantum computer, e.g.~as time evolution of coupled qubits,  or the corresponding matrix exponentiation can be computed for the simulation on a classical computer. The latter is possible for general $L\times L$ matrices using Pad\'e approximants \cite{Awad.et.al2009, Higham.Tisseur2000} in $O(L^3)$ steps. Whether the simulation of the measurement test algorithm yields an advantage over classical methods to solve linear systems of equations, which also carry complexity $O(L^3)$, depends on their accuracy and the properties of $A$.   

To the best of our knowledge, so far there is no example in the literature of a quantum algorithm to solve a linear system of equation with a dense matrix, which is presented here. This might be the case because VQLS and other methods rely on a decomposition of the observable of the cost function in terms of Pauli strings, which requires the detection of the expectation value of each of the Pauli strings separately and thus a huge overhead of runs. However, they are designed to work on NISQ computers, while the measurement test algorithm requires fault tolerance. Is it possible to replace the phase estimation algorithm inside the measurement test by a more noise robust method such as iterative or robust phase estimation \cite{Nelson.Baczewski2024}?       

The measurement test algorithm, when implemented on fault tolerant quantum computers might be applied to other computational tasks that involve the optimisation of the expectation value of a self-adjoint operator, such as the Variational Quantum Eigensolver to find ground state energies of molecules.   \\

\section*{Acknowledgements}
The authors are grateful to their collaborator E.D. Davis for in depth discussions of the material, for an analysis of the variance of the Hadamard test and  his help to determine the scaling of the accuracy.  A.G.T.\ acknowledges  funding by the South African Quantum Technology Initiative and the Department of Science, Technology and Innovation of South Africa.          
 \label{section4}
\bibliographystyle{unsrt}

\begin{thebibliography}{10}

\bibitem{HHL2009}
Aram~W. Harrow, Avinatan Hassidim, and Seth Lloyd.
\newblock Quantum algorithm for linear systems of equations.
\newblock {\em Phys. Rev. Lett.}, 103:150502, Oct 2009.

\bibitem{Rebentrost_2019}
Patrick Rebentrost, Maria Schuld, Leonard Wossnig, Francesco Petruccione, and
  Seth Lloyd.
\newblock Quantum gradient descent and newton’s method for constrained
  polynomial optimization.
\newblock {\em New Journal of Physics}, 21(7):073023, jul 2019.

\bibitem{Cerezoetal21}
M.~Cerezo, A.~Arrasmith, and R.~et~al. Babbush.
\newblock Variational quantum algorithms.
\newblock {\em Nat Rev Phys}, 3:625–644, 2021.

\bibitem{Bharti.et.al2022}
Kishor Bharti, Alba Cervera-Lierta, Thi~Ha Kyaw, Tobias Haug, Sumner
  Alperin-Lea, Abhinav Anand, Matthias Degroote, Hermanni Heimonen, Jakob~S.
  Kottmann, Tim Menke, Wai-Keong Mok, Sukin Sim, Leong-Chuan Kwek, and Al\'an
  Aspuru-Guzik.
\newblock Noisy intermediate-scale quantum algorithms.
\newblock {\em Rev. Mod. Phys.}, 94:015004, Feb 2022.

\bibitem{OBrian.et.al2014}
Alberto Peruzzo, Jarrod McClean, Peter Shadbolt, Man-Hong Yung, Xiao-Qi Zhou,
  Peter~J. Love, Alan Aspuru-Guzik, and Jeremy~L. O'Brien.
\newblock A variational eigenvalue solver on a photonic quantum processor.
\newblock {\em Nature Communications}, 5:4213, 2014.

\bibitem{Grimsley.et.al19}
Harper~R. Grimsley, E.~Economou, Sophia, Edwin Barnes, and Nicholas~J. Mayhall.
\newblock An adaptive variational algorithm for exact molecular simulations on
  a quantum computer.
\newblock {\em Nat Comm}, 10:3007, 2019.

\bibitem{Tilly.et.al22}
J.~Tilly, H.~Chen, S.~Cao, and et~al.
\newblock The variational quantum eigensolver: A review of methods and best
  practices.
\newblock {\em Physics Reports}, 986:1--128, 2022.

\bibitem{Yashioka.et.al22}
Nobuyuki Yoshioka, Takeshi Sato, Yuya~O. Nakagawa, Yu-ya Ohnishi, and Wataru
  Mizukami.
\newblock Variational quantum simulation for periodic materials.
\newblock {\em Phys. Rev. Res.}, 4:013052, Jan 2022.

\bibitem{Innan.et.al23}
N.~Innan, M.~A. Khan, and M.~Bennai.
\newblock Quantum computing for electronic structure analysis: Ground state
  energy and molecular properties calculations.
\newblock {\em Materials Today Commun}, 38:107760, 2023.

\bibitem{Li.et.al25}
X.~Li, Y.~Fan, J.~Liu, Z.~Li, and J.~Yang.
\newblock Adaptive variational quantum simulations of periodic materials using
  qubit-encoded wave functions.
\newblock {\em Journal of Chemical Theory and Computation}, 21:5973--5985,
  2025.
\newblock E-print cond-mat/0003225.

\bibitem{Romero.et.al22}
A.~M. Romero, J.~Engel, H.-L. Tang, and S.~E. Economou.
\newblock Solving nuclear structure problems with the adaptive variational
  quantum algorithm.
\newblock {\em Phys Rev C}, 105:064317, 2000.

\bibitem{Farhi.et.al2014}
Edward Farhi, Jeffrey Goldstone, and Sam Gutmann.
\newblock A quantum approximate optimization algorithm, 2014.
\newblock E-print quant-ph/1411.4028.

\bibitem{Moll.et.al2018}
Nikolaj Moll, Panagiotis Barkoutsos, Lev~S Bishop, Jerry~M Chow, Andrew Cross,
  Daniel~J Egger, Stefan Filipp, Andreas Fuhrer, Jay~M Gambetta, Marc Ganzhorn,
  Abhinav Kandala, Antonio Mezzacapo, Peter Müller, Walter Riess, Gian Salis,
  John Smolin, Ivano Tavernelli, and Kristan Temme.
\newblock Quantum optimization using variational algorithms on near-term
  quantum devices.
\newblock {\em Quantum Science and Technology}, 3(3):030503, jun 2018.

\bibitem{Diez-Valle.et.al2023}
Pablo Díez-Valle, Jorge Luis-Hita, Senaida Hernández-Santana, Fernando
  Martínez-García, Álvaro Díaz-Fernández, Eva Andrés, Juan José
  García-Ripoll, Escolástico Sánchez-Martínez, and Diego Porras.
\newblock Multiobjective variational quantum optimization for constrained
  problems: an application to cash handling.
\newblock {\em Quantum Science and Technology}, 8(4):045009, jul 2023.

\bibitem{Childs2021}
Andrew~M. Childs, Jin-Peng Liu, and Aaron Ostrander.
\newblock High-precision quantum algorithms for partial differential equations.
\newblock {\em {Quantum}}, 5:574, November 2021.

\bibitem{Balducci2022}
Giorgio Tosti~Balducci, Boyang Chen, Matthias Möller, Marc Gerritsma, and
  Roeland De~Breuker.
\newblock Review and perspectives in quantum computing for partial differential
  equations in structural mechanics.
\newblock {\em Frontiers in Mechanical Engineering}, Volume 8 - 2022, 2022.

\bibitem{Jin2023}
Shi Jin, Nana Liu, and Yue Yu.
\newblock Quantum simulation of partial differential equations: Applications
  and detailed analysis.
\newblock {\em Phys. Rev. A}, 108:032603, Sep 2023.

\bibitem{Nguyen2025}
Thanh Nguyen.
\newblock Quantum algorithms for partial differential equations: A performance
  review and future trajectories.
\newblock In Nagar Atulya~K., Jat Dharm~Singh, Mishra Durgesh~Kumar, and Amit
  Joshi, editors, {\em Intelligent Sustainable Systems}, pages 18--37, Cham,
  2025. Springer Nature Switzerland.

\bibitem{Lubasch.et.al20}
Michael Lubasch, Jaewoo Joo, Pierre Moinier, Martin Kiffner, and Dieter Jaksch.
\newblock Variational quantum algorithms for nonlinear problems.
\newblock {\em Phys. Rev. A}, 101:010301, Jan 2020.

\bibitem{Jaeger.et.al2023}
Jonas J{\"a}ger and Roman~V. Krems.
\newblock Universal expressiveness of variational quantum classifiers and
  quantum kernels for support vector machines.
\newblock {\em Nature Communications}, 14(1):576, 2023.

\bibitem{Nielsen.Chuang00}
M.A. Nielsen and I.L. Chuang.
\newblock {\em Quantum Computation and Quantum Information}.
\newblock Cambridge University Press, Cambridge and New York, 2009.

\bibitem{Bravo-Prieto23}
C.~Bravo-Prieto, R.~LaRose, M.~Cerezo, Y.~Subasi, L.~Cincio, and P.J. Coles.
\newblock Variational quantum linear solver.
\newblock {\em Quantum}, 7:1188, 2023.

\bibitem{Busch.et.al91}
P.~Busch, P.J. Lahti, and P.~Mittelstaedt.
\newblock {\em The Quantum Theory of Measurement}.
\newblock Springer-Verlag, Berlin, 1991.

\bibitem{Schuster.et.al25}
Thomas Schuster, Jonas Haferkamp, and Hsin-Yuan Huang.
\newblock Random unitaries in extremely low depth.
\newblock {\em Science}, 389(6755):92--96, 2025.

\bibitem{rotosolve}
Mateusz Ostaszewski, Edward Grant, and Marcello Benedetti.
\newblock Structure optimization for parameterized quantum circuits.
\newblock {\em {Quantum}}, 5:391, January 2021.

\bibitem{Hayashi.et.al22}
M~Hayashi, Z.W. Liu, and Yuan H.
\newblock Global heisenberg scaling in noisy and practical phase estimation.
\newblock {\em Quantum Sci.~Technol.}, 7:025030, 2022.

\bibitem{Kreyszig65}
E.~Kreysig.
\newblock {\em Statistische Methoden und ihre Anwendungen (in german)}.
\newblock Vandenhoek \& Ruprecht, Goettingen, Germany, 1970.

\bibitem{Awad.et.al2009}
A.H. .~Al-Mohy and N.J. Higham.
\newblock A new scaling and squaring algorithm for the matrix exponential.
\newblock {\em SIAM J. Matrix Anal. Appl.}, 31:970--989, 2009.

\bibitem{Higham.Tisseur2000}
N.J. Higham and Tisseur F.
\newblock A block algorithm for matrix 1-norm estimation, with an application
  to 1-norm pseudospectra.
\newblock {\em SIAM J. Matrix Anal. Appl.}, 21:025030, 2000.

\bibitem{Nelson.Baczewski2024}
Jacob~S. Nelson and Andrew~D. Baczewski.
\newblock Assessment of quantum phase estimation protocols for early
  fault-tolerant quantum computers.
\newblock {\em Phys. Rev. A}, 110:042420, Oct 2024.

\end{thebibliography}

\end{document}